\documentclass{ees2023}
\usepackage{graphicx}
\usepackage{xcolor}
\usepackage{hyperref}
\usepackage{mathrsfs}
\hypersetup{
    colorlinks = true,
    linkcolor = blue,
    anchorcolor = blue,
    citecolor = blue,
    filecolor = blue,
    urlcolor = blue
}

\usepackage[]{natbib}  

\def\BibTeX{{\rm B\kern-.05em{\sc i\kern-.025em b}\kern-.08em
    T\kern-.1667em\lower.7ex\hbox{E}\kern-.125emX}}
\bibpunct{(}{)}{;}{a}{}{,}  

\providecommand{\azero}{\ensuremath{{A_\mathrm{0}}}}

\providecommand{\ag}{\ensuremath{{A_\mathrm{G}}}}
\providecommand{\av}{\ensuremath{{A_\mathrm{V}}}}

\newcommand{\teff}[0]{\ensuremath{T_\mathrm{eff}}}

\newcommand{\logg}[0]{$\log g$}

\newcommand{\vsini}[0]{\ensuremath{\mathrm{V}\sin{i}}}
\newcommand{\kps}[0]{$\mathrm{km}\,\mathrm{s}^{\mathrm{-1}}$}

\newcommand{\grvs}[0]{\ensuremath{G_\mathrm{RVS}}}

\begin{document}

\TitreGlobal{EES 2023}


\title{Astrophysical Parameters associated to 'Hot' stars in Gaia DR3}

\author{Y. Fremat}
\address{Royal observatory of Belgium, 3 avenue circulaire, 1180 Brussels, Belgium}
\email{yves.fremat@observatory.be}





\setcounter{page}{1}


\maketitle


\begin{abstract}
Hot stars, of spectral types O-, B-, and A-, represent a small fraction of the stars observed by the Gaia satellite. Their properties and the specifications of the on-board instruments make their identification challenging. In the Gaia DR3, 12\,104\,577 targets have been assigned a temperature greater than 7500 K. It represents $\sim$2.5\% of the stars having an effective temperature value in the Gaia DR3 {\tt astrophysical\_parameters} table. We review the results obtained by the Apsis modules, focusing on the effective temperature, surface gravity, interstellar extinction, and \vsini\ parameters.
\end{abstract}



\section{Introduction}

In this contribution, we deal with and refer to 'hot' stars, defined as stars with an effective temperature (\teff) greater than 7500 K and of spectral types O, B, or A (OBA). They are located on the upper main sequence, beyond the onset \teff\ of strong convection, and are, on average, rotating significantly faster than cooler stars. For this reason, they are good laboratories to study stellar evolution at high angular rates. Usually more massive than two solar masses, hot stars also have a higher luminosity and can be observed from large distances. They belong to young stellar populations 
often embedded in open clusters and star forming regions found in the galactic disc, which makes their light highly absorbed by interstellar gas and dust (i.e., ISM, interstellar medium). All along their evolution, they interact dynamically and chemically with the surrounding ISM, ionising its particles and enriching its chemical composition. OBA stars are important to identify among the huge amount of Gaia data, as they are a key to a better understanding of the galaxy's chemical content, structure, evolution, and dynamics.

Their Gaia spectra are dominated by broad hydrogen lines (Balmer series in the BP/RP and Paschen series in the RVS), by the near-IR calcium triplet lines (in the RVS), and by the Balmer jump lying at the very blue end of the BP passband. These features, however, decrease very rapidly with effective temperature, leaving an almost featureless spectral energy distribution (SED, see Figs.\,4 and 6 of \citealt{2023A&A...674A..28F}) and making their analysis challenging \citep[e.g.,][]{2023A&A...674A...7B}. The paper focuses on the astrophysical parameters (APs) of hot stars available in the Gaia DR3 {\tt astrophysical\_parameters} table \citep{2023A&A...674A..26C}. We provide a rapid overview of their behaviour and limits of applicability.

\section{Classification of hot stars by the Apsis pipeline}

Two methods are providing astrophysical parameters for stars with $\teff > 7500$~K: the General Stellar Parametri\-zer from Photometry (GSP-Phot) and the Extended Stellar Parametrizer for Hot Stars (ESP-HS). Both modules make different assumptions and analyse epoch-combined spectroscopic data (internally calibrated BP/RP and RVS). While GSP-Phot is aimed at treating all the stars, whatever their effective temperature, ESP-HS focuses specifically on the treatment of hot stars. Their $G$ magnitude coverage is also different \citep[see Fig.~2 of ][]{2023A&A...674A..26C}; the former is treating stars brighter than $G = 19$, the latter those with $G \le 17.65$.

GSP-Phot uses a Markov Chain Monte Carlo (MCMC) approach to derive the \teff, surface gravity (\logg), metallicity ([M/H]), and interstellar extinction parameters (\azero, \ag, ...). Its results are based on the analysis of the sampled BP/RP spectra, the apparent $G$ magnitude, and the parallax, which are confronted with simulated spectra, photometry, and isochrones. GSP-Phot considers four different synthetic spectral libraries individually (i.e., APs per library are saved in table {\tt astrophysical\_parameters\_supp}), then determines which one provides the best fit (i.e., best fit APs are saved in table {\tt astrophysical\_parameters}, with field names having the suffix {\tt \_gspphot}). More information about the algorithms is available in the dedicated paper of \citet{2023A&A...674A..27A}.

The ESP-HS algorithm is based on a minimum distance approach to derive \teff, \logg, \vsini\ (when RVS data of sufficient quality is available), and \azero\ (i.e., APs are saved in table {\tt astrophysical\_parameters}, with field names having the suffix {\tt \_esphs}). In the first step, it uses a simplex downhill algorithm to fit BP/RP and, when available, RVS simulated spectra to the observations. The distinction between these two processing modes is made in the first digit of the flag {\tt esphs\_flags}: 0 (BP/RP+RVS fit) and 1 (BP/RP-only fit). The algorithm assumes a solar chemical composition, keeping [M/H] at zero. In the process, both passbands are normalised separately. In a second step, a Levenberg-Marquadt algorithm is applied to fine-tune the solution and derive the uncertainties. The synthetic spectra used to fit the BP/RP observations are taken from a library computed with the {\tt LLmodels} code \citep[7500 to 20\,000 K: ][]{2004A&A...428..993S} and, for the hotter stars, with the {\tt synspec/tlusty} computer programme \citep{1988CoPhC..52..103H} adopting the BSTAR \citep{2007ApJS..169...83L} and OSTAR \citep{2003ApJS..146..417L} atmosphere models. The corresponding simulations were then generated with the SMSGen tool \citep[Sampled Mean Spectrum Generator; see Sect. 2.3.2 in][]{2023A&A...674A..26C}. Systematic mismatches found in the comparison of these simulations and observations for a sample of B and A-type stars with known APs derived from Stromgren photometry led to the masking of the spectrum above 800\,nm and to the semi-empirical correction of the BP/RP data (mainly below 400\,nm). The synthetic spectra used to model the RVS data were convolved with a Gaussian spectroscopic line spread function (LSF) and by assuming a median resolving power of $\mathscr{R} = 11\,500$ \citep[e.g.][]{2023A&A...674A...5K}. In practice, the co-added RVS spectra analysed by the Apsis modules are a mixture of epoch spectra, each having their own LSF. This LSF is known to be non-Gaussian and may have $\mathscr{R}$ values significantly different from the documented median. It was the intention to use the fitting of the rotational broadening parameter to compensate for any deviation from the adopted assumption.

\begin{figure}[!htpb]
\centerline{\includegraphics[width=14cm]{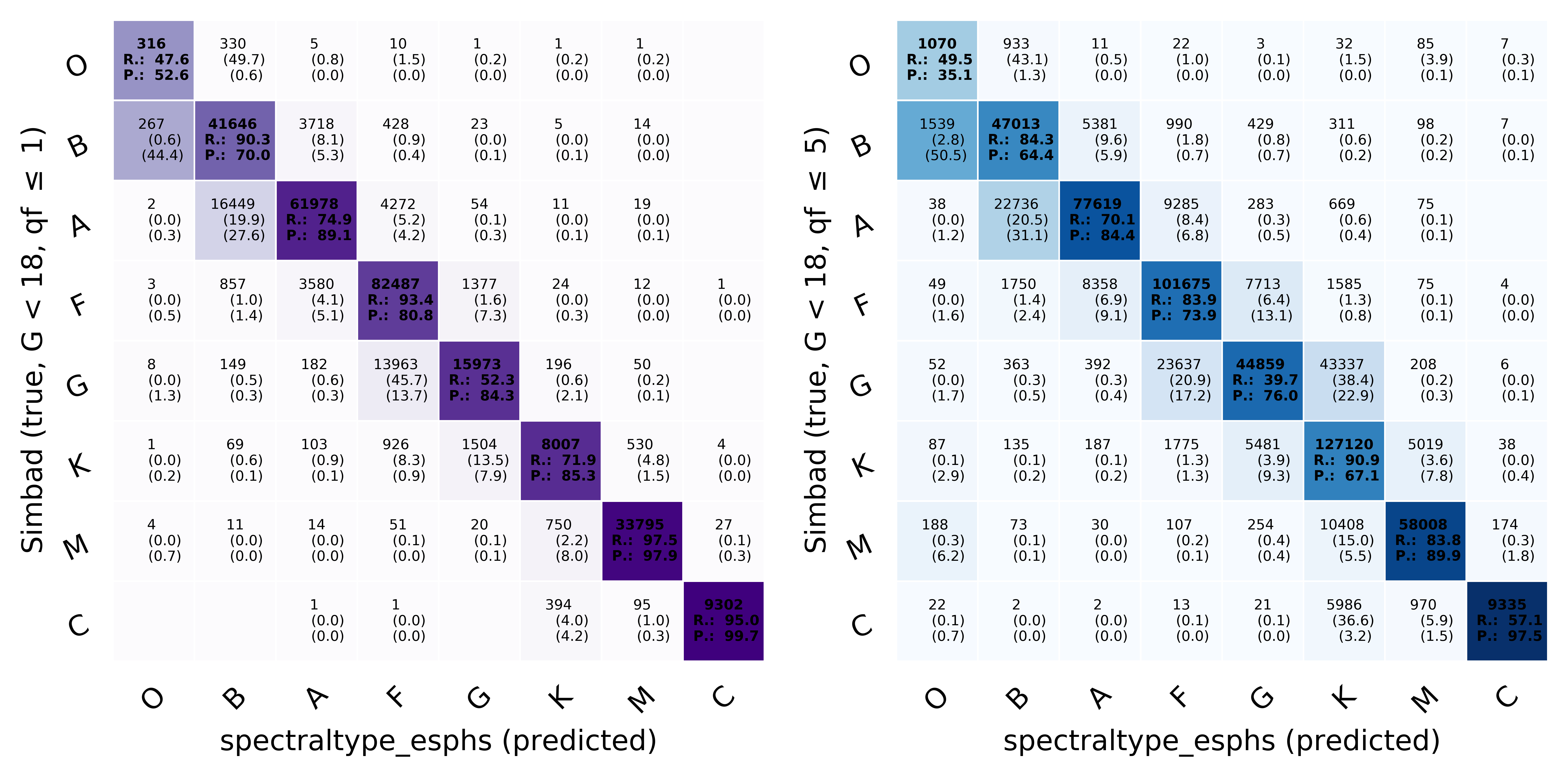}}
\caption{Confusion matrix of {\tt spectraltype\_esphs} in Gaia DR3 (abscissa) with the spectral type label (ordinate) available in the Simbad database (CDS). The top number in each cell gives the number of targets of a given Simbad spectral type that received a certain predicted tag. The second and third numbers give the corresponding percentage along the row and column, respectively. The colour scale follows the column percentage (i.e., the third number). The right panel takes all the targets into account, whatever the quality flag (qf) value is. In the left panel, we only consider those stars classified with the highest probability.\label{fig:matrix}}
\end{figure}

Unlike GSP-Phot, ESP-HS is not applied to all the data. It is launched on the O, B, and A-type stars flagged by a random forest algorithm trained on a data sample made of simulated and observed \citep[see Appendix A.4. of][]{2023A&A...674A..26C} BP/RP spectra. Because it is used to pre-select the hot stars, we found it useful to save the spectral type tag in the Gaia DR3 catalogue \citep[field {\tt spectraltype\_esphs}, whose definition and temperature dependance are given in Table A.1. of][]{2023A&A...674A..26C}. 
We show in Fig.\,\ref{fig:matrix} the confusion matrix obtained by correlating the {\tt spectraltype\_esphs} tag to the spectral type information found in Simbad at the CDS. Because the temperature coverage of the various spectral types may differ, confusion between neighbouring types is expected. Globally, we obtain a fair level of consistency between both classifications. The O-type star classification is usually less accurate. These are often confused with B-type stars, and, at $G$ magnitudes fainter than 15, a significant fraction of M stars are tagged as (highly reddened) O-type stars. A quality assessment is also provided with the tag. It is saved as the second digit of the field {\tt esphs\_flags} (i.e., {\tt esphs\_flags}[1:2], with integer values ranging from 1 to 5). The lower the value of this quality flag, the closer the observed spectrum is expected to match the training sample. Finally, except for the candidate Be stars identified by the ESP-ELS algorithm, ESP-HS is not processing targets with significant H$\alpha$ line emissions.

\section{Astrophysical parameters}

GSP-Phot and ESP-HS provided for Gaia DR3 an estimate of the astrophysical parameters (i.e., $\teff > 7500$K) of 11\,156\,494 and 2\,382\,015 targets, respectively. The overlap between the two samples comprises 1\,433\,932 objects. The different biases, magnitude domains, and post-processing criteria explain why the ESP-HS sample is not completely included in GSP-Phot.

\begin{figure}[!htbp]
\centerline{\includegraphics[width=0.9\textwidth]{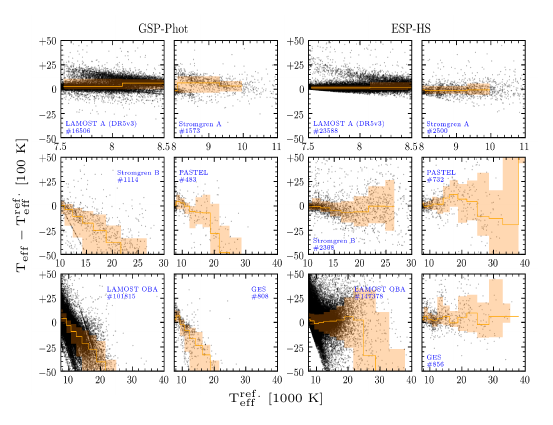}}
\caption{Effective temperature residuals obtained by GSP-Phot (left panel) and ESP-HS (right panel) relative to various reference catalogues are plotted against the reference effective temperature. Considered catalogues and surveys are: LAMOST A catalogue \citep[][DR5v3]{2019yCat.5164....0L}, Stromgren \citep[derived within DPAC adopting the updated calibration of] []{1993A&A...268..653N}, PASTEL \citep{2016A&A...591A.118S}, LAMOST OBA \citep[][DR6]{2022A&A...662A..66X}, Gaia ESO Survey \citep[GES: ][]{2022A&A...661A.120B}.\label{fig:tempres}}
\end{figure}

\begin{figure}[!htbp]
\centerline{\includegraphics[width=13cm]{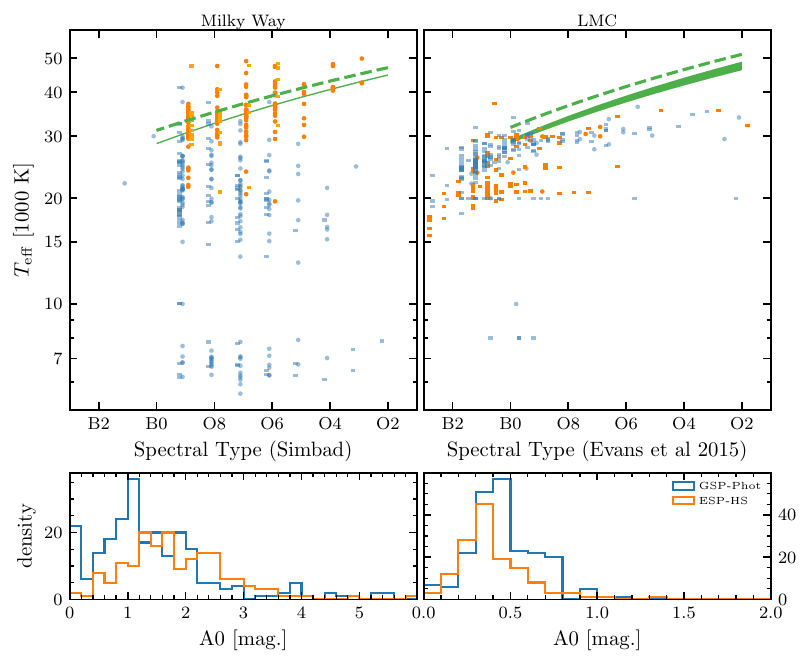}}
\caption{{\bf Upper panels:} Effective temperature derived by GSP-Phot (blue symbols) and ESP-HS (orange symbols) plotted as a function of the spectral type found in Simbad (left panel for galactic stars) and published by \citet{2015A&A...584A...5E} (LMC stars in right panel). The spectral type vs. \teff\ calibrations found by \citet{2010A&A...524A..98W} for Milky Way and LMC dwarfs (broken green curve) and giants (solid green curve) are compared to the Apsis measurements. {\bf Lower panels:} distribution of the \azero\ extinction in the galaxy (left) and LMC (right) derived by the two Apsis modules.\label{fig:tempscale}}
\end{figure}

In order to assess their accuracy and precision, the \teff\, and \logg\, results were compared with those found in various catalogues. These catalogues were chosen for having, when combined, temperature and magnitude domains matching those of the Apsis' OBA sample. Figure\,\ref{fig:tempres} shows how the \teff\ residuals (i.e., measured minus expected value) vary with the catalogue \teff\, estimate. In the A-type star domain, the two modules provide consistent \teff\, measurements with similar, usually positive, biases (from -50 to 300 K) and spread (200 to 600 K). The sequence of positive outlying residuals
seen in the 'LAMOST A' panels of Fig.\,\ref{fig:tempres} might be related to the combined effect of the parameter accuracy and of the hydrogen lines having their maximum intensity around 8500 K \citep{2022A&A...662A..66X}. At hotter effective temperatures, ESP-HS measurements tend to be underestimated by a few thousand K above 25\,000 K while the scatter also increases with the effective temperature (from $\sim$2000 K to $\sim$7000 K for the early O-type stars). The temperatures provided by GSP-Phot tend to deviate systematically above 10\,000 K with negative biases. Therefore, it can be considered less reliable than ESP-HS for B-type stars. For O-type stars, we may have a better idea of the trends by comparing the APs to the Galactic and LMC spectral types vs. temperature scale (Fig.\,\ref{fig:tempscale}, upper panels). In our galaxy, ESP-HS usually derives a larger interstellar extinction (A0, lower panels of Fig.\,\ref{fig:tempscale}) and \teff, that better matches the calibration determined by \citet{2010A&A...524A..98W}. GSP-Phot obtains systematically lower estimates due to non-adapted priors and a difficulty in selecting the 'best library' fit. For Large Magellanic Cloud (LMC) stars, in which direction the interstellar extinction is lower, the conclusions are reversed. GSP-Phot tends to provide a temperature scale closer to the one of \citet{2010A&A...524A..98W} for the LMC.

\begin{figure}[!htbp]
\centerline{\includegraphics[width=0.9\textwidth]{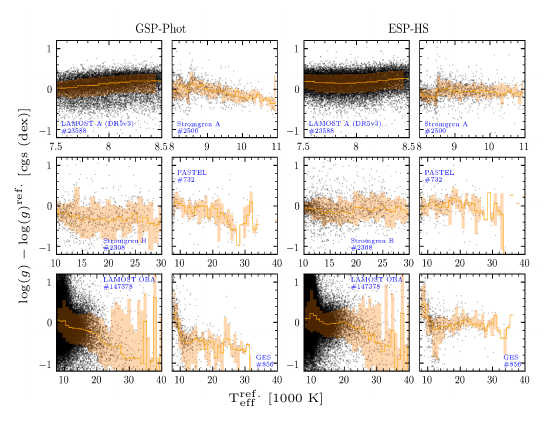}}
\caption{Same as Fig.\,\ref{fig:tempres} for the surface gravity.\label{fig:loggres}}
\end{figure}

The distribution of the surface gravity residuals with \teff\, is shown in Fig.\,\ref{fig:loggres}. The \logg\, offsets differ from one module to the other, usually by 0.1 dex in the 7500 - 10\,000 K temperature domain, with ESP-HS providing larger estimates. When comparing the results to the GES survey, the behaviour of the residuals significantly and consistently changes around 10\,000 K in the two datasets. This behaviour is also seen in the comparison with benchmark stars in \citet[Fig.\,3,][]{2022A&A...661A.120B}. As explained by the same authors and also discussed by \citet{2022A&A...662A..66X}, it could be due to the Balmer lines reaching their maximum strength around 8500~K. Above this \teff, GSP-Phot results are impacted by the offsets found in temperature and are usually underestimated. On the other hand, ESP-HS \logg\ measurements show a constant bias that varies from $-$0.1 to $+$0.1 dex, mainly depending on the origin of the comparison values. The scatter is of the order of 0.2 to 0.4 dex in the considered magnitude range.

\begin{figure}[!htbp]
\centerline{\includegraphics[width=\textwidth]{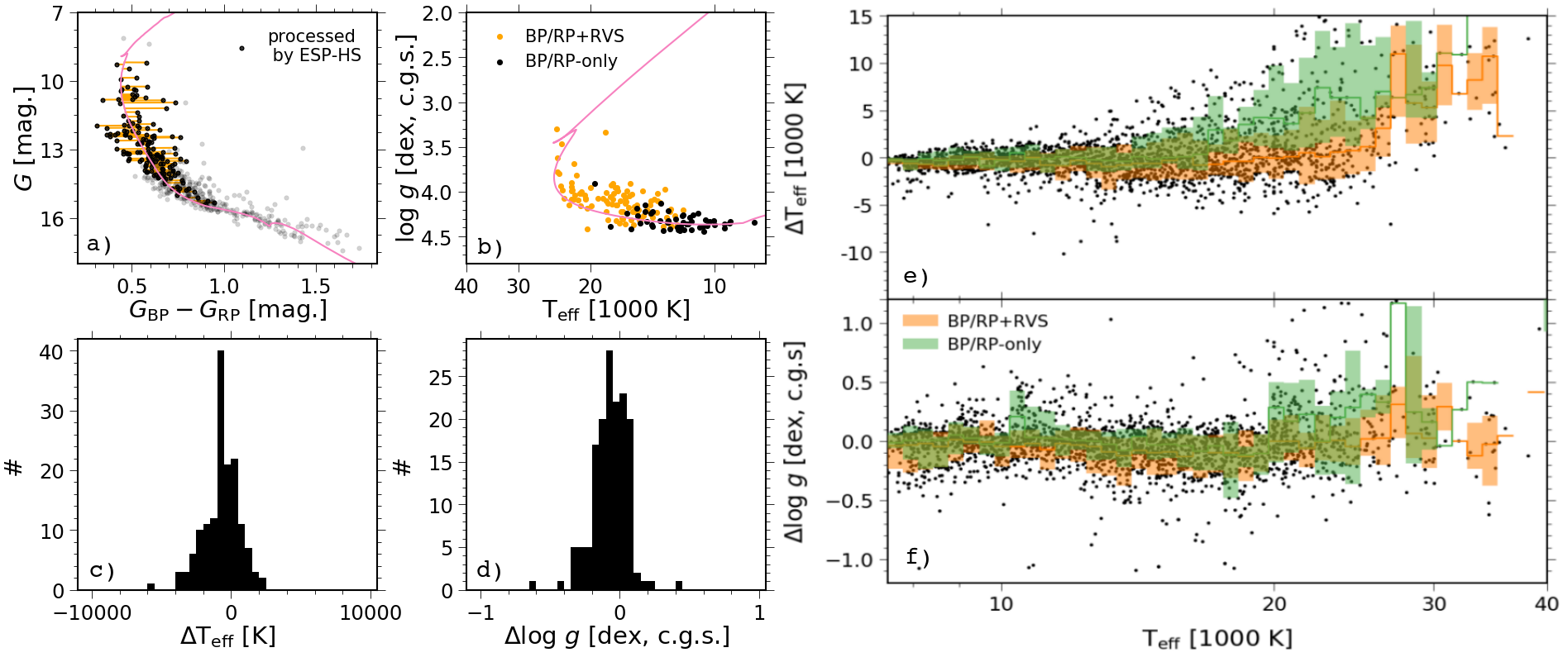}}
\caption{The colour diagram (Panel a) of the cluster NGC869 is used to identify the closest point in the corresponding reddened PARSEC isochrone (pink curve; Nb of targets with APs: 152, $\log(t)$ = 7.18, $\av=1.68$). The grey points are targets without any published ESP-HS AP. The corresponding Kiel diagram is shown in Panel b by adopting the APs derived by ESP-HS and by making the distinction between the two processing modes (i.e., BP/RP+RVS or BP/RP-only fit). The isochrone \teff\, and \logg\, are then used to compute the residuals as shown in Panels c) and d). The same approach was used for 1524 members of 42 open clusters to analyse the offset and spread of ESP-HS APs (Panels e and f). The running median is represented by the solid line step function, while the interquantile range is represented by shades. Different colours are used to make the distinction between the two processing modes.\label{fig:clusters}}
\end{figure}

ESP-HS results were also confronted with the reddened PARSEC\footnote{PARSEC isochrones (v1.2S and CMD 3.6 web interface) are available from \url{http://stev.oapd.inaf.
it/cgi-bin/cmd}} isochrones \citep{2012MNRAS.427..127B, 2015MNRAS.452.1068C} of open clusters. In Fig.\,\ref{fig:clusters}, the membership, extinction and ages, t, are taken from \citet{2020A&A...640A...1C}. The effective temperature offset (in BP/RP+RVS mode) is of the order of $-$300~K for A-type stars, $+$400~K for B-type stars, and reaching $+$6000~K in the O-type stars' \teff\ range. The \logg\ median bias is of the order of $-$0.02 to $-$0.08 dex. It is usually negative and constant with temperature. Compared to the spectroscopic ground-based estimates, the \teff\ offset has a different sign. The difference due to the processing mode becomes significant above 10\,000 K, where the BP/RP-only ({\tt flags\_esphs[0:1]} = 1) processing produces a somewhat larger \teff\ offset ($+$1200 K in the 20000 - 30000 K domain) than the simultaneous fitting of BP/RP and RVS ($+$700 K in the 20000 - 30000 K domain). The dispersion remains similar in both cases.

From the comparison of the Apsis results to the temperature scale found in the MW and LMC O-type stars (Fig.\,\ref{fig:tempscale}), the degeneracy between temperature and interstellar extinction tends to lead to an underestimate of the effective temperature in both modules. However, while GSP-Phot and ESP-HS \azero\ interstellar extinction estimates are not identical, they remain consistent. Differences become more obvious at $\teff>$10\,000~K \citep[Fig.\,14 in][]{2023A&A...674A..26C}, where ESP-HS results tend to be slightly larger because of the different treatment of the BP/RP (i.e., the module is ignoring the near-IR and applying an empirical correction to the simulated spectra). When the results obtained in open clusters are considered, the two algorithms consistently provide larger monochromatic extinctions \citep[Fig.\,30 in][]{2023A&A...674A..28F} than the values found in \citet{2020A&A...640A...1C}. 

\begin{figure}[!htbp]
\centerline{\includegraphics[width=0.9\textwidth]{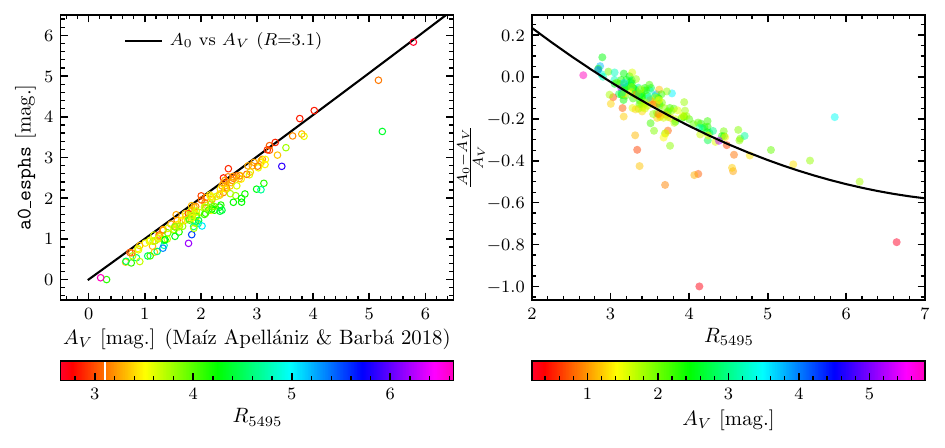}}
\caption{{\bf Left panel:} Comparison of the \azero\ monochromatic extinction value obtained by ESP-HS for a sample of galactic O-type stars to the \av\ found in \citet{2018A&A...613A...9M}. The black line represents the relation that links \azero\ to \av, when R = 3.1. The colour coding follows the ratio of total to selective extinction, $R_\mathrm{5495}$, found by the same authors (see colour bar, where the horizontal white line represents the $R$=3.1 value assumed in Apsis modules). {\bf Right panel:} We plot the relative extinction difference as a function of $R_\mathrm{5495}$ (black dots). The blue curve shows a second-degree polynomial through the data. The colour scale follows the \av\ \citep{2018A&A...613A...9M}.\label{fig:azero}}
\end{figure}

To generate the BP/RP simulations that the algorithms confront to the observations, the Apsis team chose the interstellar extinction function derived by \citet{1999PASP..111...63F} and fixed the ratio of total to selective extinction at $R=3.1$. Especially for hot stars, which are often found towards highly reddened lines of sight, these assumptions may have a significant impact \citep[e.g.,][]{2024arXiv240101116M}. They lead to spectra vs. template mismatches, can make the \teff\, vs. \azero\ degeneracy complexier, and bias the \azero\ value itself. For ESP-HS, we show in Fig.\,\ref{fig:azero} the order of magnitude of the effects on \azero\ and compare its results to those obtained by \citet{2018A&A...613A...9M} who determine both Av and $R$ (i.e., $R_\mathrm{5495}$ in their paper) simultaneously. The relative difference between the two estimates is usually negative, \azero\ being smaller, and its absolute value increases with $R$.

\begin{figure}[!htbp]
\centerline{\includegraphics[width=0.6\textwidth]{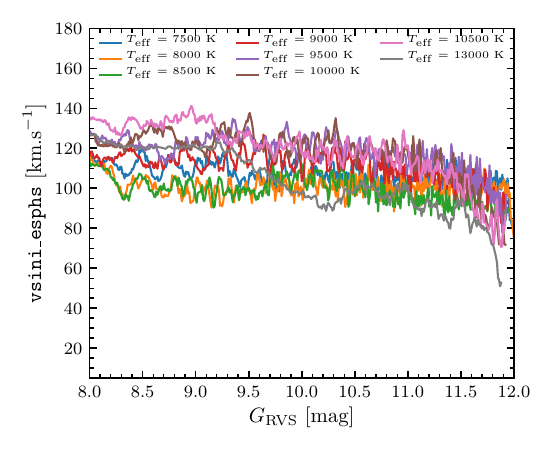}}
\caption{Running median of {\tt vsini\_esphs} as a function of the \grvs\ magnitude. Various temperature bins ({\tt teff\_esphs $\pm$ 250~K}) are considered and represented by different colours.\label{fig:vsinimag}}
\end{figure}

Two estimates of the rotational broadening are available in the Gaia DR3 catalogue: {\tt vbroad} in table {\tt gaia\_source} and {\tt vsini\_esphs} in {\tt astrophysical\_parameters}. The former was determined by the CU6 pipeline on the per transit (i.e., epoch spectrum) RVS spectrum for all stars cooler than 15\,000 K; the latter was measured simultaneously with the APs on the co-added RVS of stars hotter than 7500 K. The behaviour of {\tt vbroad} is described in \citet{2023A&A...674A...8F}.
A comparison between the two measurements is shown in Fig.\,11.82 of the online Gaia DR3 documentation (Sect.\,11.4.4). At lower \vsini, ESP-HS estimates tend to be larger than {\tt vbroad}. It is due to the accuracy of the radial velocities being lower in hot stars \citep{2023A&A...674A...7B} and to the assumptions made on the LSF. The median {\tt vsini\_esphs} of the stars with $\teff > 7500$~K is found to be relatively constant with \grvs, and ranges from 100 to 140~\kps\ as also found for main sequence stars in the \vsini\ compilation of \citet{2001yCat.3226....0G}. At $\grvs > 11.5$, the median is dropping rapidly, which puts a magnitude threshold above which the \vsini\ derived from co-added RVS spectra should be disregarded as less accurate. Fig.\,\ref{fig:vsinimag} can directly be compared to Fig.\,14 of \citet{2023A&A...674A...8F}, which is its equivalent obtained for {\tt vbroad}. As expected, the upper magnitude threshold of {\tt vbroad} is, for the hot stars, lower and therefore less robust with \grvs\ than the value obtained from co-added spectra. At temperatures greater than 10\,000~K, the main criteria remaining to derive the projected rotational velocity are the broad Paschen lines. These lines are very sensitive to temperature and surface gravity variations, and therefore any offset on the APs directly affects the accuracy of {\tt vsini\_esphs}. For this reason, the results obtained above 25\,000 K must be considered with caution, especially at $\grvs > 10$.

\begin{figure}[!htbp]
\centerline{\includegraphics[width=0.8\textwidth]{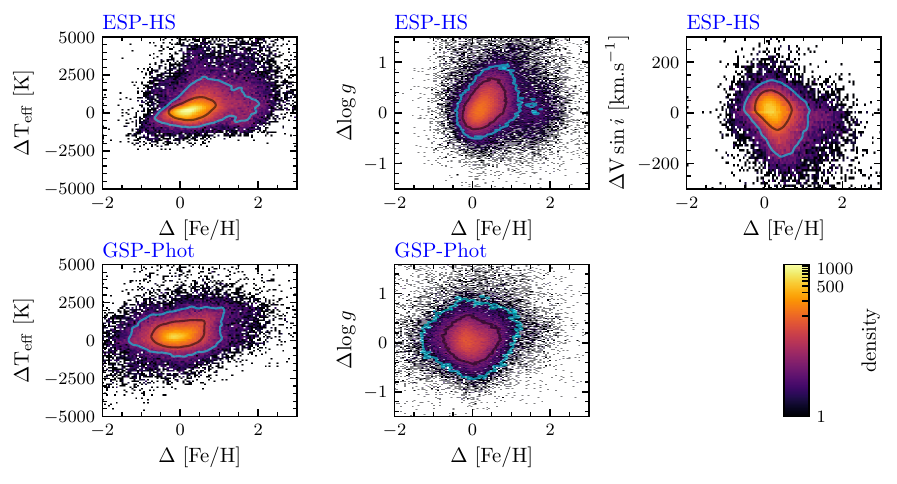}}
\caption{Effective temperature, surface gravity, and \vsini\ residuals \citep[i.e., Apsis value minus the one derived from LAMOST spectra by][]{2022A&A...662A..66X} are plotted as a function of the [Fe/H] residuals. As GSP-Phot is most reliable below 10\,000~K, we limited the comparison to A-type stars. We defined the black and cyan contour lines such that they encompass 68\% and 90\% of the targets, respectively.
Upper panels: correlation of the residuals obtained by ESP-HS. Lower panels: correlation of the residuals obtained by GSP-Phot.
\label{fig:fecov}}
\end{figure}

In Fig.\,\ref{fig:fecov}, we explore the impact of having metallicity mismatches on the APs computed by ESP-HS assuming a solar chemical composition. For the sample studied by \citet{2022A&A...662A..66X}, we compare the AP residuals to the metallicity mismatches (upper panels of Fig.\,\ref{fig:fecov}). The plots are limited to stars with $\teff \le $10\,000~K, so that we can confront these density maps to those obtained for the same targets by GSP-Phot (lower panels of Fig.\,\ref{fig:fecov}). The surface gravity and the \vsini\ residuals are the most correlated with [Fe/H] mismatches. 

\begin{figure}[!htbp]
\centerline{\includegraphics[width=0.7\textwidth]{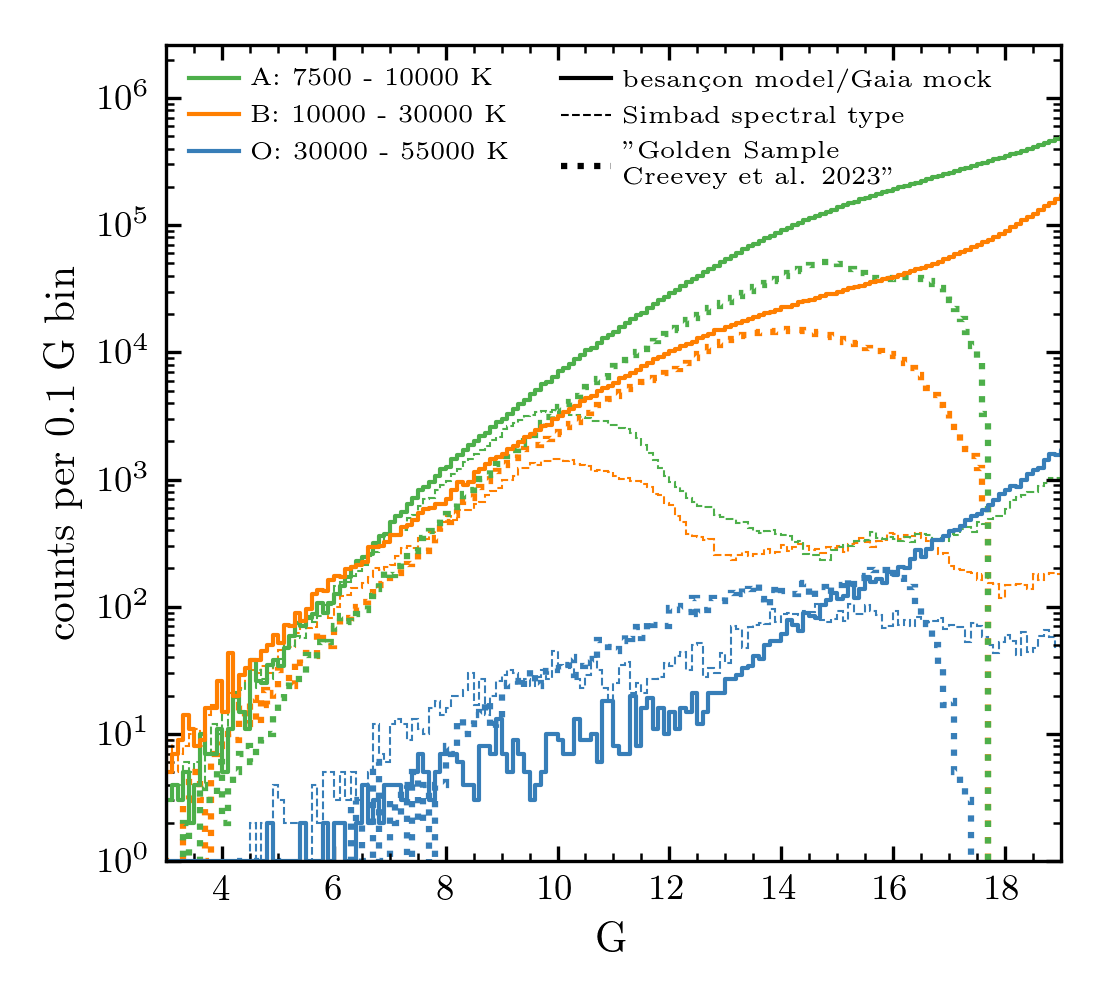}}
\caption{Magnitude distribution of A-, B-, and O-type stars. The thick coloured lines represent the distribution as found in the GeDR3mock catalogue \citep{2020PASP..132g4501R,2012A&A...543A.100R}. The spectral type definition is given in the legend. The thin broken lines show the counting of hot stars done in 2022 using the basic information tab of the Simbad database. The magnitude distributions of the OBA 'golden sample' \citep{2023A&A...674A..39G}
is plotted with thick dotted lines.
\label{fig:magdist}}
\end{figure}

\section{Discussion and conclusions}

In Fig.~\ref{fig:magdist}, we present the magnitude distribution of A-, B-, and O-type stars as it is expected by the Besan\c{c}on
galactic model and the GeDR3mock catalogue \citep{2020PASP..132g4501R,2012A&A...543A.100R}. We also compare these distributions to those found in the Simbad database in 2022. These statistics do not take into account the most recent classification work done in various ground-based surveys, but one may still expect to identify with Gaia a significant number of hot stars at magnitudes $G > 12$.
The difficulty is knowing how representative of the galactic hot star population such a sample is and how pure and complete it can be. 

A significant effort was made all along in the validation of the results to remove from the catalogue spurious AP estimates. The criteria used during the post-processing were often based on the offline analysis of the goodness-of-fit distributions. However, the candidate list of O-, B-, and A-type stars is still polluted by lower-mass stars. In the list, we still find white dwarfs (WDs), subdwarf stars, and hot horizontal branch stars (hot HBs) that share the same effective temperature as more massive hot stars. The presence of hot HBs, for example, is seen through overdensities in the Kiel diagram of stars with magnitude $G > 14$ \citep[right panel of Fig.\,19 in][]{2023A&A...674A..26C}, while WDs are better seen in the absolute magnitude vs. \teff\ diagram. More unexpected is the presence of cooler stars, such as RR-Lyrae or LPVs, in the sample of hot stars. These misclassifications can be attributed to the combination of AP degeneracy (i.e., effective temperature and interstellar extinction), template mismatches (our simulations do not reproduce all the existing classes of stars and chemical compositions), and stellar variability (e.g., the number of combined epochs is often different in the BP and RP passbands). 

To improve the cleaning of the hot star sample, other criteria were applied to the available APs after the post-processing. These are described in \citet{2023A&A...674A..39G} and resulted in the publication with Gaia DR3 of a list of 3\,023\,388  candidate OBA stars ({\tt gold\_sample\_oba\_stars}). Their magnitude distribution is plotted in Fig.\,\ref{fig:magdist}. For the A- and B-type stars, it fits well with the expectations of the GeDR3mock catalogue up to magnitude 14. The drawback of certain filtering criteria is the removal of bona fide OBA stars on the bright and faint ends of the magnitude domain. The impact of the filters and of the remaining \teff\ vs. interstellar extinction degeneracy on the completeness of the sample is shown in Fig.\,2 of \citet{2023A&A...674A..39G}, where the fraction of identified to expected OBA stars in open clusters is reported as a function of \azero. The higher \azero, the more hot stars are missing.

\begin{acknowledgements}
We thank the BELgian federal Science Policy Office (BELSPO)
for funding us through the PROgramme de Développement d’Expériences
scientifiques (PRODEX) and the grant 4000140452. This research has made use of the SIMBAD database, operated at CDS (Centre de Donn\'ees astronomiques de Strasbourg), Strasbourg, France.
\end{acknowledgements}

\bibliographystyle{aa}  
\bibliography{ees2023-yfremat} 

\end{document}